# Agent Based Processing of Global Evaluation Function

MAHMUD SHAHRIAR HOSSAIN
*Lecturer, Department of Computer Science and Engineering*
*Shahjalal University of Science and Technology, Sylhet-3114, Bangladesh.*
*E-mail: shahriar-cse@sust.edu*

M. MUZTABA FUAD
*Department of Computer Science*
*Montana State University, Bozeman, MT 59717, USA.*
*E-mail: fuad@cs.montana.edu*

DR. MD. MAHBUBUL ALAM JOARDER
*Assistant Professor, Institute of Information Technology (IIT)*
*University of Dhaka, Dhaka-1000, Bangladesh.*
*E-mail: joarder@udhaka.net*
***FAX: +880-2-8615583***

*Load balancing across a networked environment is a monotonous job. Moreover, if the job to be distributed is a constraint satisfying one, the distribution of load demands core intelligence. This paper proposes parallel processing through Global Evaluation Function by means of randomly initialized agents for solving Constraint Satisfaction Problems. A potential issue about the number of agents in a machine under the invocation of distribution is discussed here for securing the maximum benefit from Global Evaluation and parallel processing. The proposed system is compared with typical solution that shows an exclusive outcome supporting the nobility of parallel implementation of Global Evaluation Function with certain number of agents in each invoked machine.*

***Keywords:*** Global Evaluation Function, Constraint Satisfaction Problem, Backtracking and Homogenization.

## 1. INTRODUCTION AND OVERVIEW

Problem solving is a very important topic in Artificial Intelligence. Decision Problems or Constraint Satisfaction Problems (CSP)[1] are critical because some of them need the best solution while others may not have any solution at all. Typical solutions like backtracking cannot solve such problems because for large-scale problems, searching space would increase sharply. There are several basic algorithms and methods for solving Constraint Satisfaction Problems. Single-solution algorithms harness their computing power for only one solution in a time. Examples can be provided as Backtracking[4], Extremal Optimization[6], Local Search[7], Simulated Annealing[5], Alife&AER[2, 3] model, etc. These algorithms can be added to multi-agent system. Jing et al.[1] only focus on such one-copy algorithms with an emergence of Local Evaluation Function. Our research work is based on Agent based Global Evaluation Function, while it focuses on concurrent placement of agents at different locations for its computation power.

Generate-test (GT)[1] generates a possible combination of all variables and then checks whether it satisfies all the constraints or not. The simplest way to generate a complete assignment of all variables is to select a value randomly, for each variable. However this is a very inefficient way. GT can grow much smarter with the application of sharp intelligence so that the random generation of test cases is forcefully headed toward the solution. This can be done by an evaluation mechanism of the current situation of the problem-state. This evaluation mechanism can be based on Global Evaluation Function (GEF), which would vary problem to problem, depending on the constraints to be satisfied. Once again, to avoid falling into the local-optima, the algorithm may sometimes perform stop-and-restart[8], random walk and Tabu search[9, 10] .

If multiple agents are created to perform GEF based GT, the forceful run toward the solution would be faster. If the number of agents in a standalone machine is increased, where all the agents are initialized randomly, the searching capability would grow larger. But after including certain number of agents, the system performance would decrease although it seems that the searching capability is higher. The included agents with random initialization will not be able to secure the benefit any more because the speed of each agent toward the solution would decrease due to increasing number of agents in one machine. So the number of agents in a standalone machine should be minimized by rigorous analysis.

If a distributed system is used for GEF based GT, each machine should run multiple agents to secure the maximum benefit from the system. But the number of agents in a machine should be optimized. This paper focuses on parallel agent based solution by GEF for harnessing maximum processing power gained from a distributed system.

## 2. ALGORITHM AND STRATEGY

This paper organizes relationship between Constraint Satisfaction Problem, Global Evaluation Function and Multi-agent systems. Further it goes forward to show its distributed implications.

### 2.1 CONSTRAINT SATISFACTION PROBLEM (CSP)

A Constraint Satisfaction Problem, ***P***, consists of[11]

1. A finite set of variables, ***X***, identified by the natural numbers, *1, 2, ………, n*, $\boldsymbol{X}=\{X_1,X_2,\dots\dots\dots,X_n\}$.
2. A domain set, containing a finite and discrete domain for each variable. $\boldsymbol{D}=\{D_1,D_2,\dots.D_n\}$, for all $i \in [1, 2, \dots\dots., n]$, $X_i \in D_i$ .
3. A constraint set, $\boldsymbol{C} = \{C(R_1), C(R_2),\dots\dots\dots,C(R_m)\}$. $R_i$ is the *i*th constraint. Each $R_i$ is an ordered subset of the variables, and each constraint $C(R_i)$ is a set of tuples indicating the mutually consistent values for the variables in $R_i$.
4. *P* is a finite discrete CSP, if all $D_i$ in ***D*** are discrete and finite.
5. *P* is a binary CSP, if each constraint is either unary or binary. It is possible to convert a CSP with *n-ary* constraints to another equivalent binary CSP[11].

A solution, ***S***, is a complete assignment to all variables such that the assignment satisfies all the constraints.

In this paper, N-queen problem is chosen as a Constraint Satisfaction Problem where N queens should be placed in an N*N chessboard so that no two queens are placed in the same row, same column or same diagonal position. It is a classical benchmark of CSPs. Solutions exist for N-queen problem with N is greater than or equal to 4. The equivalent CSP is

$\boldsymbol{X} = \{ X_1, X_2, \ldots\ldots\ldots, X_n\}$ where $X_i$ refers to the position of the queen in the $i$th row.

$\boldsymbol{D}=\{D_1,D_2,\ldots.D_n\},\ \forall i,\ D_i = [1, n]$

$\boldsymbol{C}=\{C(R_u) \mid \forall i,\ j \in [1,n],\ C(R_u)=\{<b,c> \mid b \in D_i,\ c \in D_j,\ b \neq c,\ i-j \neq b-c,\ i-j \neq c-b\}\}$

N-queen problem is an attractive candidate for algorithm performance evaluation and also it is a scalable problem that has many applications[12].

## 2.2 GLOBAL EVALUATION FUNCTION (GEF)

The Global Evaluation Function for Generate-test style search methods are as follows[1]:

1. Pick a random state from possible state set.
2. Choose a neighboring state and the system moves to that state:
   Compute the evaluation value of all neighbors of the current system state by Global Evaluation Function and then select the best state; Alternatively, perform some random walk with some probability.
3. Repeat 2. and 3. until a solution is found or reached the maximum tries.
4. Return the current state as the solution.

There are varieties of neighborhood structure. One simple method to construct the neighbor is to change any one variable's assigned value. Therefore, the difference between the current state and its neighbors is one variable's assigned value, which changes the evaluation worth derived from Global Evaluation Function. By this way a number of $\sum_i \left( |D_i| - 1 \right)$ neighbors are found for each state.

GEF evaluates how good the current system-state $S_{sys}$ is. This function is used to rank the current state and to find out the best neighboring state by a move. There are lots of ways to define the evaluation function. Evaluation function plays a dominating role in the algorithm and a good definition of it will increase the performance of the algorithm[1]. It is called Global Evaluation Function because it considers the whole system, in together.

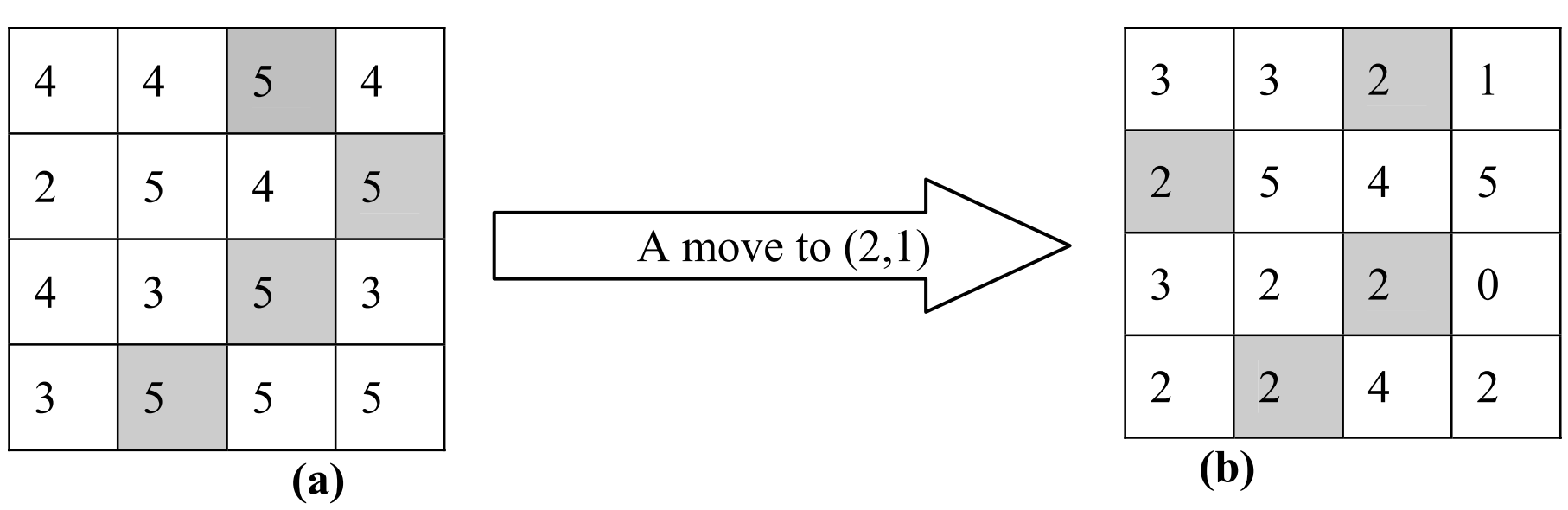

**Figure 1:** A move toward minimum evaluation value.

For N-Queen problem GEF can be defined as the number of dissatisfied constraints that follows Equation (1).

$$E_{global}(S_{sys}) = \frac{1}{2}\sum_{i=1}^{N}\sum_{j=1}^{N}\left[(s_i = s_j) \vee (s_i - s_j = i - j) \vee (s_i - s_j = j - i)\right] \quad (1)$$

The number of dissatisfied constraints in a solution state is $E_{global}$*(Solution State) = 0.* Figure 1(a)[1] shows an instance where the gray cells indicate the positions of queens for a 4-Queen problem. The number in each cell indicates the value derived by GEF for corresponding placement of queens. The current state corresponds to an evaluation value of 5 where it suggests that if the queen of the second row were moved to the first column, the evaluation value would become 2, the minimum among all other moves. Thereafter Figure 1(b) [1] shows that after the move, a solution state with evaluation value 0 is found in (3, 4). At this moment, if the queen of the third row is moved to fourth column, a solution is found.

A state with better evaluation value does not always mean that it is closer to the solution. Moreover a better evaluation value can be found in more than one cell. If the so-called better evaluation value does not provide further improvement, another cell with same evaluation value can be selected. Eventually, if all the cells with the minimum evaluation value fall under the same fallacy, the move can be given to previous one that forked the minimum value and move can be given to such a cell that have minimum evaluation value which is greater than the previous minimum.

### 2.3 AGENT-BASED EVALUATION STRATEGY

Jing et al.[1] describes agent-based solution where each variable $X_i$ is handled by one agent. This conversion of CSP to multi-agent system complicates the GEF strategy. Moreover agents cannot be controlled for global evaluation if they are placed in independent threads. Even if piping mechanism were used for inter-thread communication, the overhead would lead a dominating part. Moreover a distributed system will suffer from communication overhead if agents are placed in different machines.

This paper proposes placement of agents in independent threads so that the threads can be placed in different machines under a Local Area Network. Each agent would search the solution independently using GEF. A closer observation in the performance analysis section will show that the number of agents in one machine demands some analysis for securing the maximum benefit. The initiator of the searching should follow the following strategy:

1. Start all the agents concurrently
    a. Each agent picks a random state from its possible state set.
    b. Choose a neighboring state and the system moves to that state:
       Compute the evaluation value of all neighbors of the current system state by Global Evaluation Function and then select the best state; Alternatively, move toward better position with some probability.
    c. Repeat b. and c. until a solution is found.
    d. Return the current state to the initiator as the solution.
2. If a solution is found, notify all agents to stop their corresponding searching.

As all the agents start searching from random instances, it is possible that some of them start with lower evaluation value than the others.

## 3. PERFORMANCE ANALYSIS FOR AGENT BASED EVALUATION

The performance analysis is done in several steps to understand the behavior of the problem and it is compared with typical backtracking algorithm to solve the N-queen problem. The analysis finally harnesses the strategy to find out the maximum number of agents, *maxAgent* that can be placed in one machine to secure the benefit of random initialization of all the agents.

### 3.1 BACKTRACKING vs. GLOBAL EVALUATION FUNCTION (GEF)

Figure 2 shows a comparison between two standalone versions of solution of N-queen problem. The red line stands for the backtracking solution and the other for the GEF solution. The initial positions of all the queens are chosen randomly for GEF method. Thereafter it goes forward to the positions with lower evaluation value. It stops its searching until it finds the lowest possible evaluation value, $E_{min}$. The value $E_{min}$ is zero for the N-queen problem. Whenever it finds a position with $E_{global} = E_{min}$, it is guaranteed that it has found a solution because global evaluation value $E_{min}$ derives state with fulfillment of constraints. Figure 2 shows time to reach a solution for different number of queens. Definitely, GEF method shows better performance than typical backtracking method.

Once again, Figure 3 shows plot for a fixed number of queens, which depicts that the deterministic approach of backtracking method would result in static time for each request. The red straight line stands for typical backtracking algorithm. Alternatively, the same experiment with GEF method results in varying time because of its random initialization every time a request is made. Among all the requests 70% requests result in less time than typical backtracking algorithm. This behavior is really a promising one.

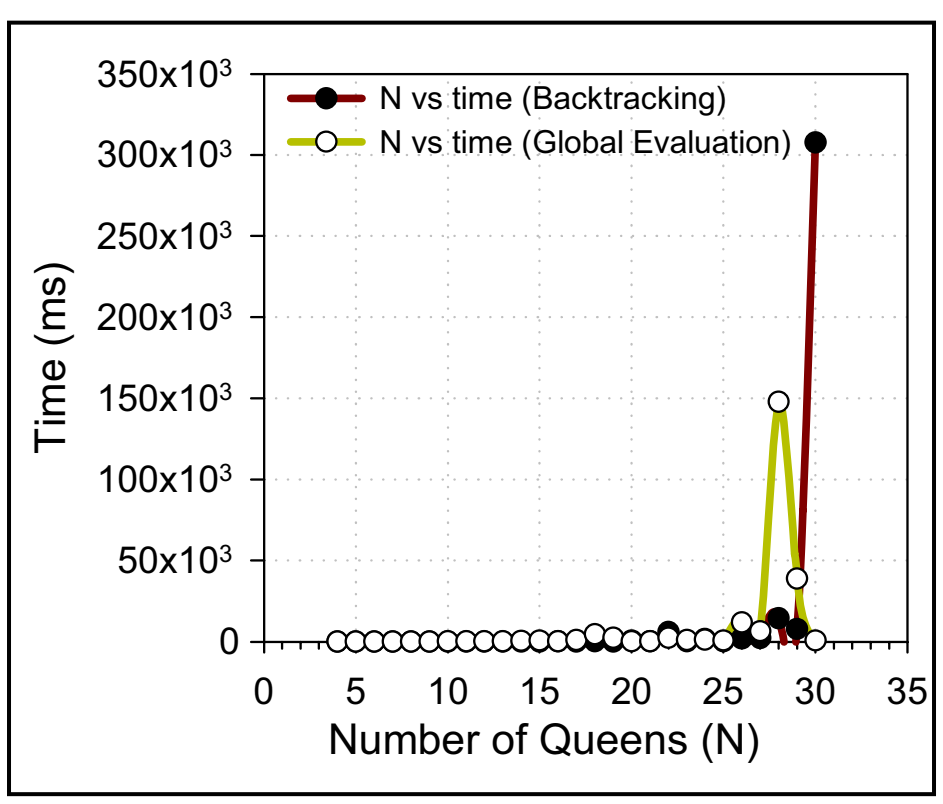

**Figure 2:** Comparison between backtracking and global evaluation function method.

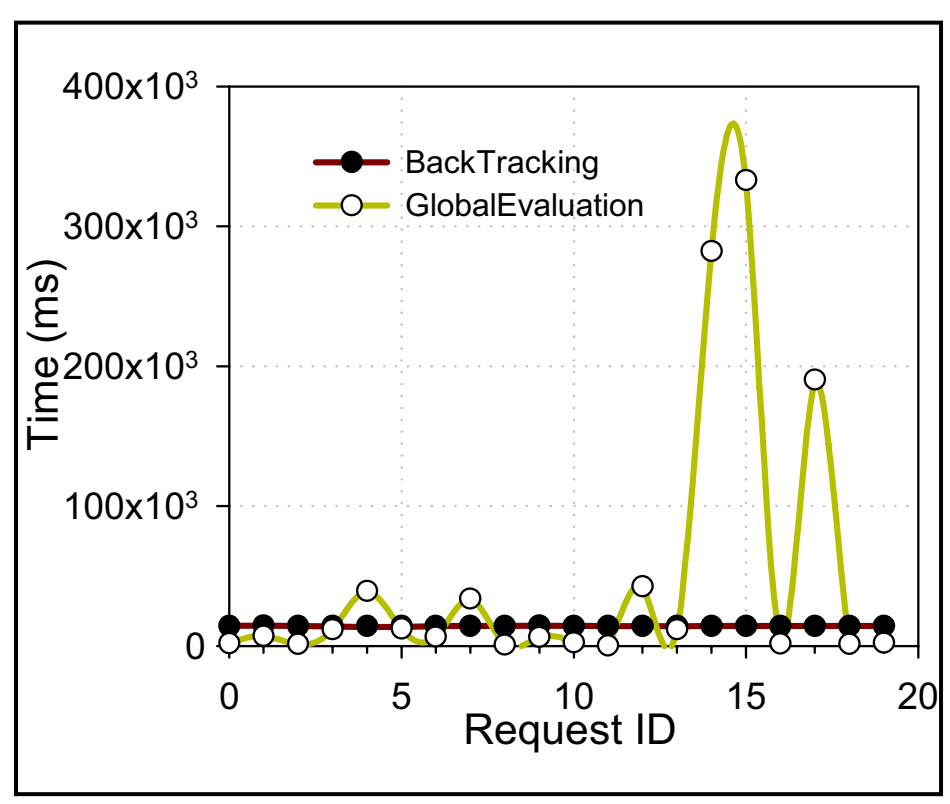

**Figure 3:** Comparison between backtracking and GEF method for ***N***=28.

Because it shows, if agents are placed in different machines it is probable that 70% machines would provide faster results than typical backtracking solution. So definitely, the initiator would get the solution faster than a typical backtracking as well as single GEF method.

### 3.2 GEF AND AGENT-BASED SOLUTION

A system that runs multiple agents concurrently in a standalone machine by different independent threads, would result in a system state that possesses the probable lower evaluation value and likely to be closer to the solution. In this way, agent-based GEF provides exclusive outcome.

Figure 4 shows the experimental results. Performances are taken for various numbers of queens, *N*. For each *N*, time to find the solution is taken, with different number of agents. A maximum of 20 agents are taken for each *N*. The experiment shows that it is better to use two agents rather than using one, again it is better to use three agents rather than two. Further inference of the statement is not true i.e., after introduction of the third agent, the search time eventually increases with the involvement of further agents. The tendency is a valid one because at every level, requests are made several times and average time is taken to plot the graph of Figure 4.

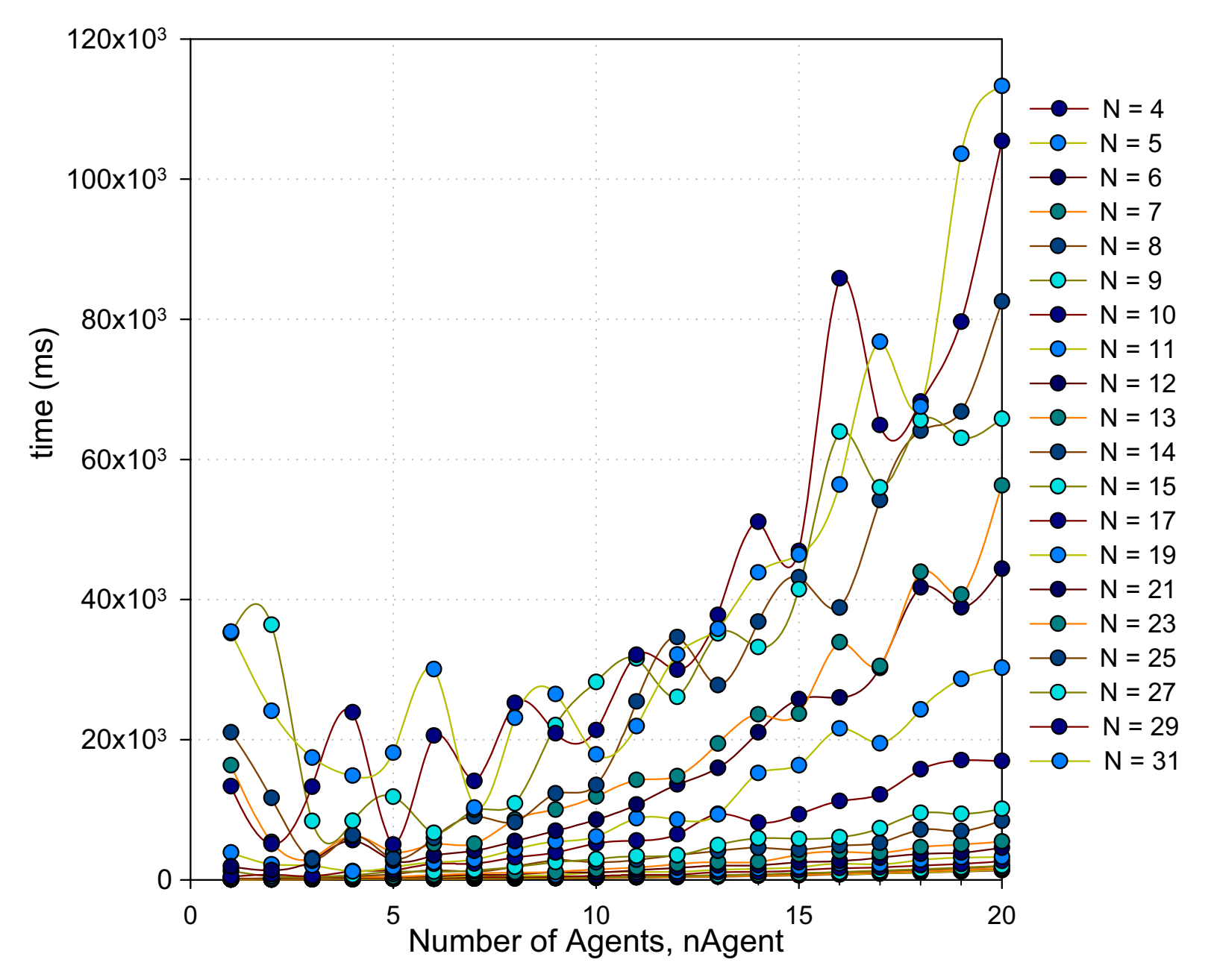

**Figure 4:** GEF for different number of queens (nAgent vs. time).

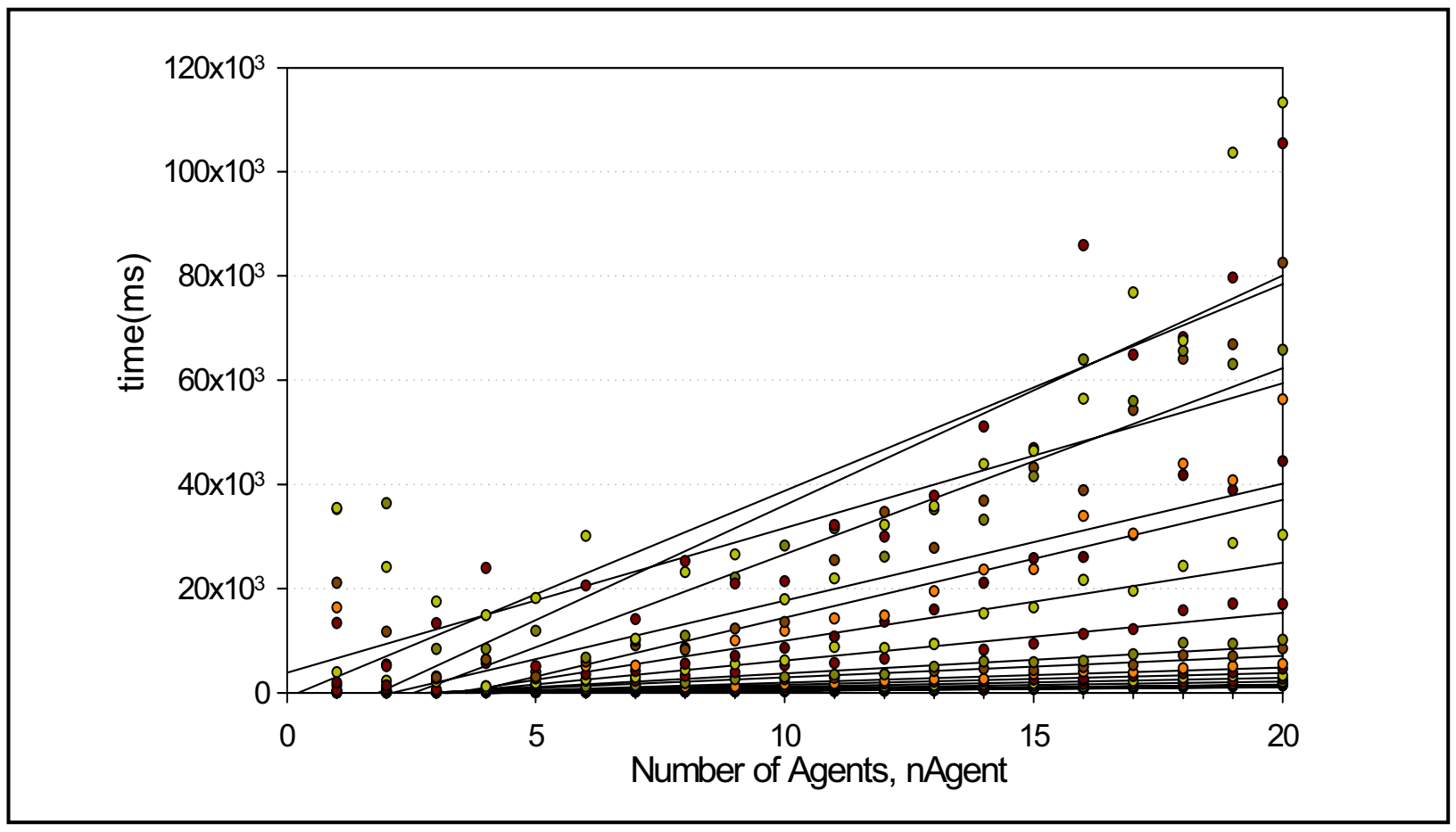

**Figure 5:** GEF for different number of queens and their tendency.

The tendency for each *N* can be retrieved from the corresponding regression lines drawn in Figure 5. Most of the regression lines crisscross *time=0* line when number of agents is between 2 and 3. The regression lines prove that if the number of agents were three, the system would provide solution in the fastest possible speed. This performance was taken in a machine with 1.7 GHz Genuine Intel processor and 256 MB of RAM.

A single regression line can be plotted by the data from which averages were taken. This regression line will provide the maximum number of agents, *maxAgent*, that should be used in a standalone machine. Figure 6 provides the scatter plot of all the data and the corresponding regression line [green line] that crisscrosses *time=0* line at nAgent = 3.

Thus, random initialization of agents can secure a benefit form the GEF but the benefit is limited in standalone machine. The number is limited by *maxAgent, k* [value of *k* depends upon some other parameters related to the performance and configuration of the machine. For our case *k*=3].

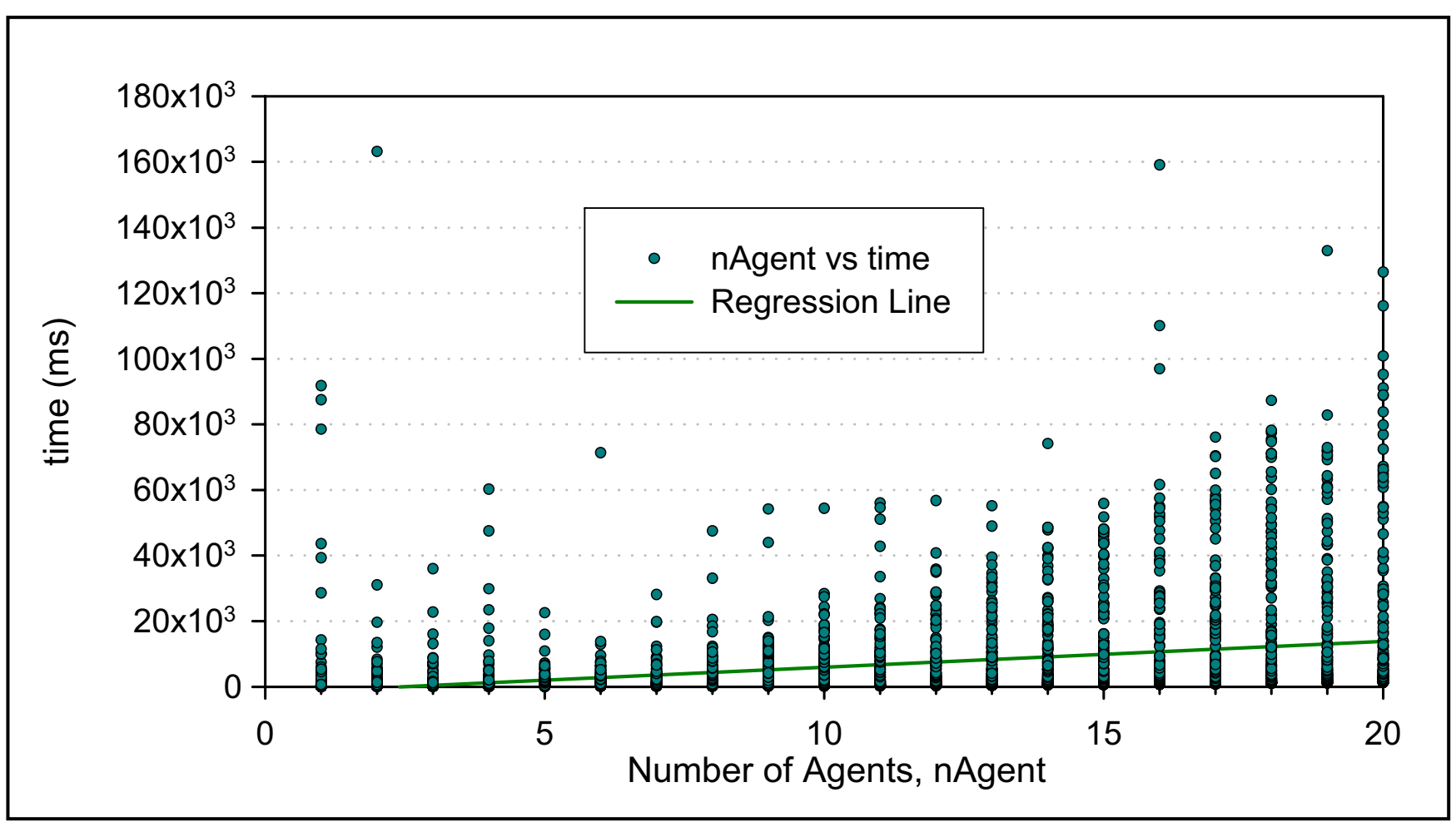

**Figure 6:** Scatter plot of all the data and the regression line.

### 3.3 HOMOGENIZATION[13] FOR *maxAgent*

Homogenization is a technique that enables the distribution of workload to different nodes in a Local Area Network composed of heterogeneous elements. Homogenization is especially applicable for problem instances that are linearly divisible. It is not applicable for dividing a problem instance of AI where probabilistic Global Evaluation mechanism is used. Despite homogenization shows its benefit in such problems by varying the number of agents in the machines. Analysis of the previous section shows that *maxAgent* is the number of agents that secures the maximum benefit of Global Evaluation Function from a machine. This number varies machine to machine depending on the performances. In the parallel implementation of Global Evaluation Function, homogenization can be used to balance the number of agents depending on the performances of the machines.

In the experiments different machines with different loads with different performances were taken and it is found that *maxAgent* varies with the machine performance. A machine performance is taken by key encryption capability of the machine per second. It is found that *maxAgent* follows equation (2).

$$maxAgent = \frac{7000 - e^{p}}{1000} \tag{2}$$

where $p$ = performance of the machine in hundred thousand keys per second.

Figure 7 shows runtime behavior and the behavior observed from the formula. It shows that as the performance is increased *maxAgent* falls accordingly. For a very high performance machine the theoretical *maxAgent* would become less than unity. In such a case, for such machines number of agents should be equal to unity. The agent loader of a machine should load $k$ number of agents where $k$ is equal to *maxAgent* from equation (2).

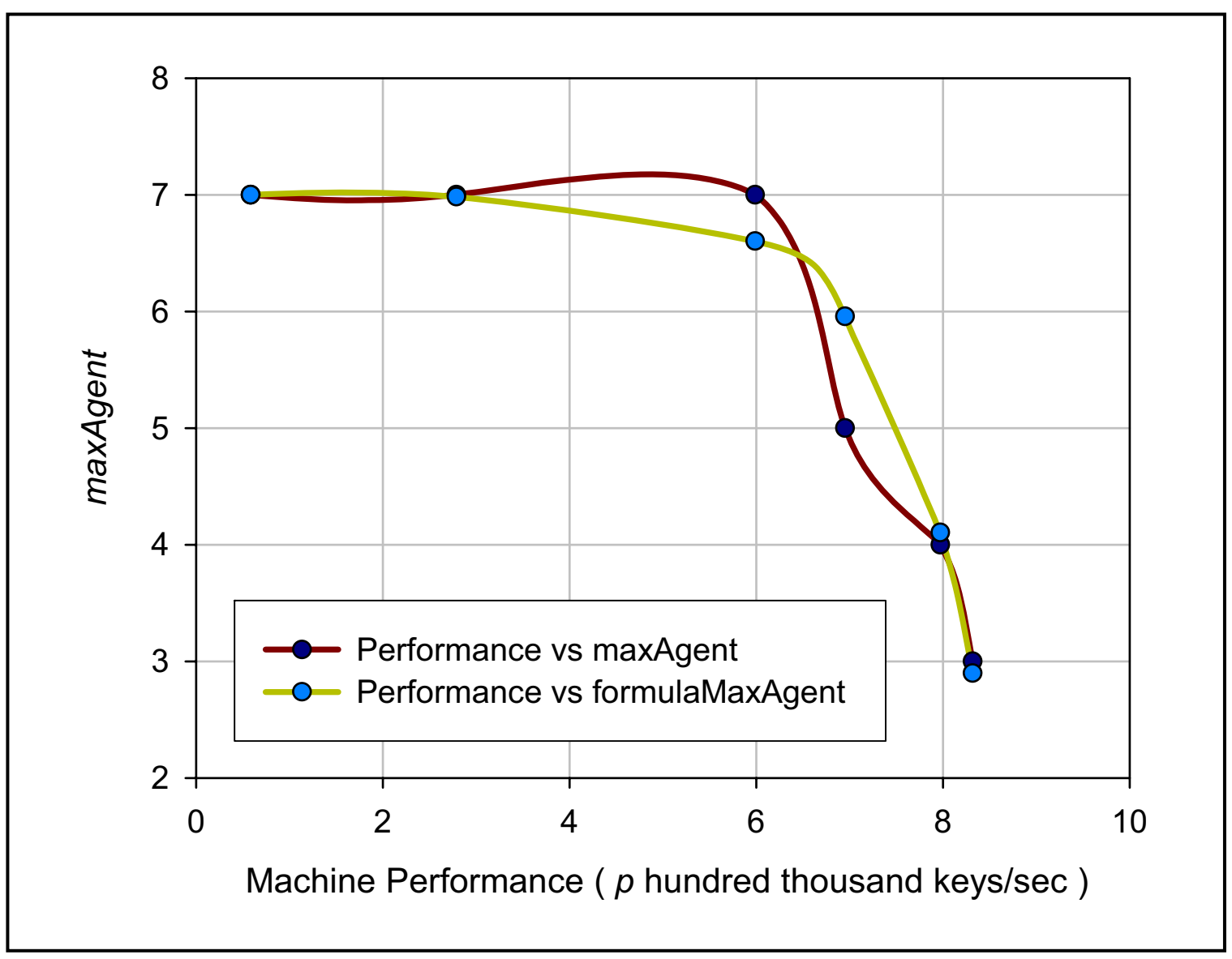

**Figure 7:** Homogenization pattern for *maxAgent*.

### 3.4 PARALLEL AGENT-BASED SOLUTION

For achieving the maximum benefit of randomization from a distributed architecture, each machine should possess a number of $k$ agents that is equal to *maxAgent*. When multiple agents are placed on each machine of a local area network, the system would return the solution within minimum possible time. By the consequence, it would suffer from an overhead $T_{OVE}$ that results from three basic areas: distribution of job, redirection of result and time to stop agents after a result is found. The *Turn around Time*, $T_{TAT}$ is the sum of *Actual Search Time*, $T_O$ and the overhead time, $T_{OVE}$ i.e.,

$$T_{TAT} = T_O + T_{OVE} \tag{3}$$

The search space increases if the number of queens, $N$ is increased. The search space increases by the square of $N$ in the N-Queen problem. As a result,

$$T_O \propto N^2 \tag{4}$$

Once again, if the number of agents, $N_A$ is increased, *Actual Search Time,* $T_O$ decreases because of the parallel randomization of agents from different locations of the local area network. *Actual Search Time*, $T_0$ is inversely proportional to the number of agents, $N_A$. That is,

$$T_O \propto \frac{1}{N_A} \tag{5}$$

Combining equation (4) and (5)

$$T_O \propto \frac{N^2}{N_A} \tag{6}$$

Once again, from equation (6) it is obtained that

$$T_O = \frac{K_O N^2}{N_A} \tag{7}$$

where $K_O$ is the constant of proportionality. The behaviors for different number of agents are reflected in Figure 8. The corresponding runtime plot is provided with Figure 9. Both the figures show that as the number of queens increases, the *Actual Search Time* increases. Moreover the lines depict that increase of number of agents results in a faster solution. Each line of Figure 8 actually reflects the behavior of equation (4) although the overall plot is done by using equation (7). The plot of Figure 8 appears in a different manner in Figure 10. Figure 11 shows its corresponding runtime plot. Both Figure 10 and Figure 11 reflect the behavior of equation (5).

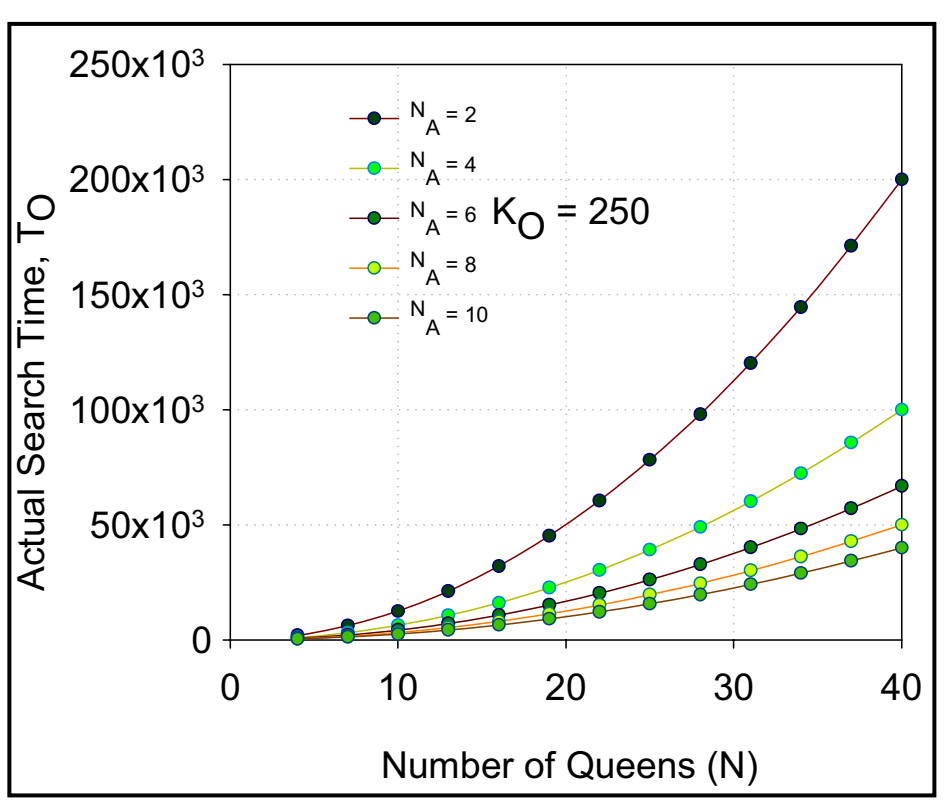

**Figure 8:** Tendency of *Actual Search time* when ***N*** increases. Different lines show the behavior with different number of agents.

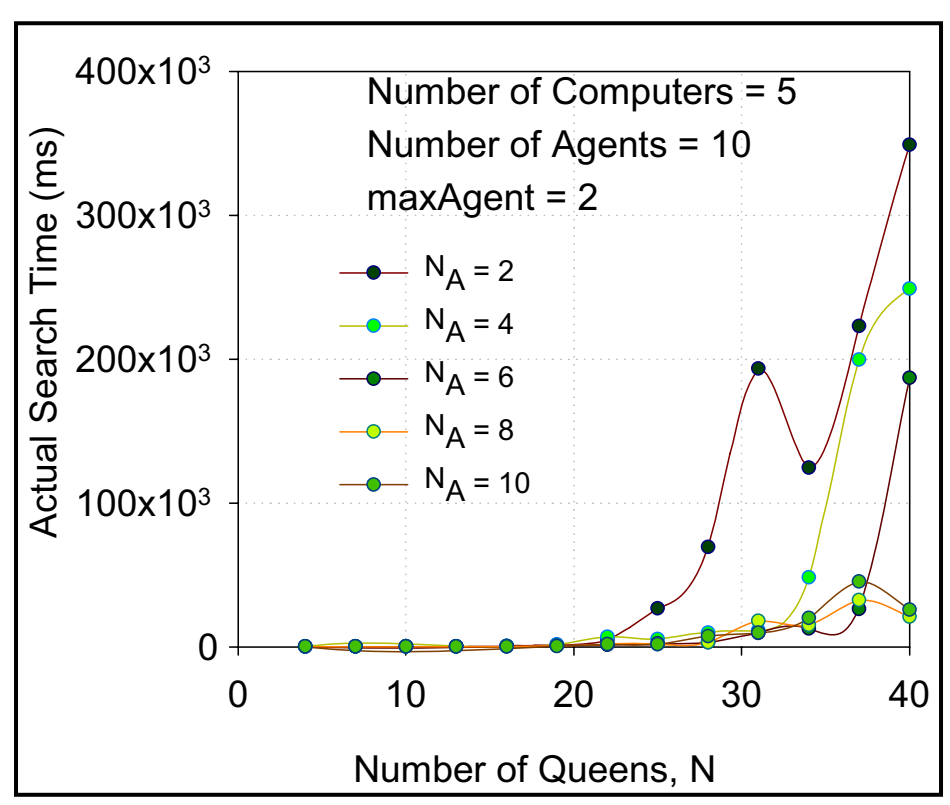

**Figure 9:** Runtime behavior of *Actual Search Time* when N increases. Different lines depict the behavior with different number of agents.

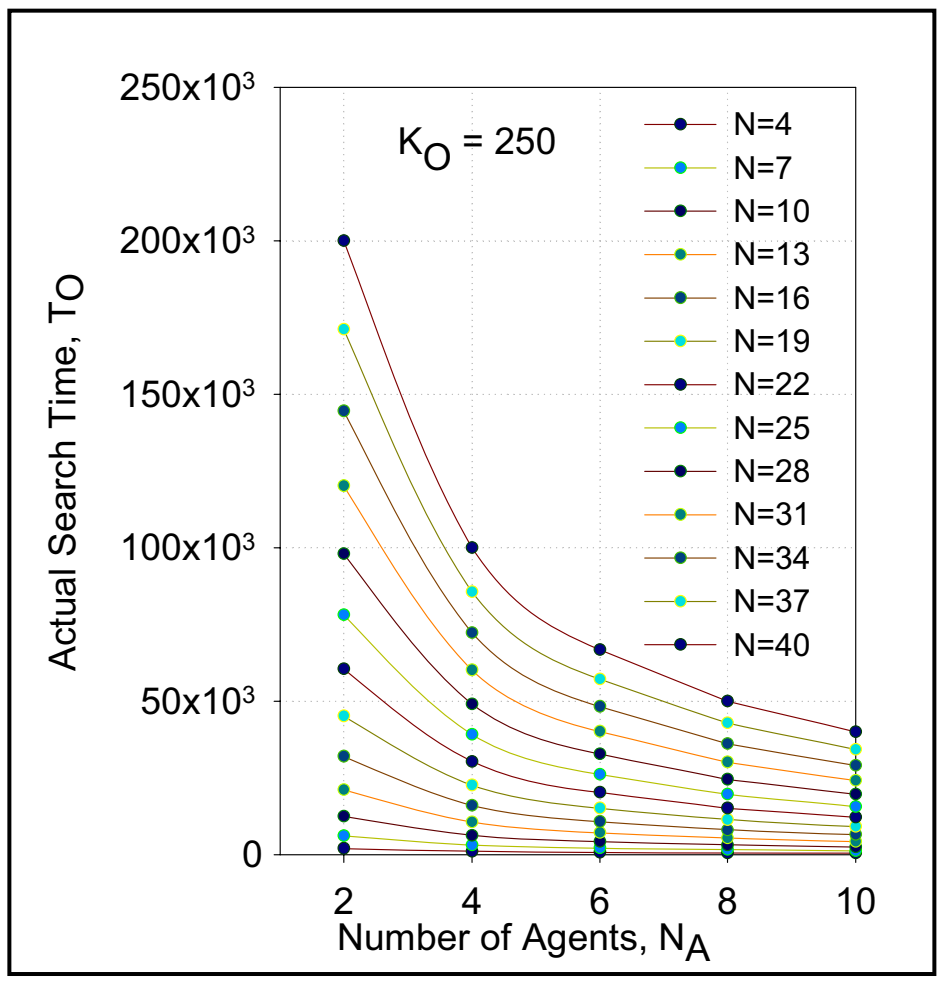

**Figure 10:** Tendency of Actual Search time when $N_A$ increases. Different lines show the behavior with different number of queens.

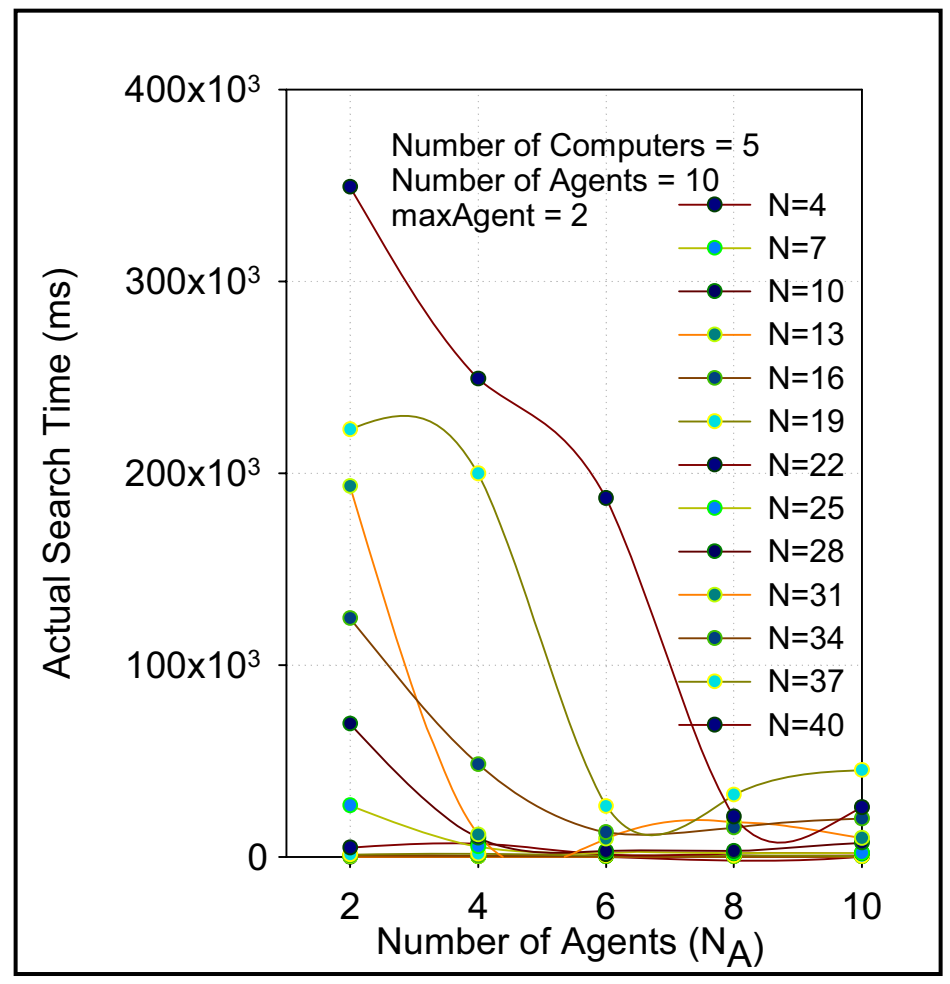

**Figure 11:** Tendency of Actual Search time when $N_A$ increases. Different lines show the behavior with different number of queens.

All the runtime simulations in this section are taken with several homogeneous machines, each with Genuine Intel 1.1 GHz processor and 128 MB of RAM. All the machines run in the Windows platform and are equipped with Java Virtual Machine provided by Sun's J2SDK 1.4.1. 100 Mbps Ethernet Network connects all the machines.

The distribution penalty, $T_{OVE}$ is proportional to the number of agents, $N_A$, because the initiator initiates all the $N_A$ agents and as well it stops all of them after receiving the result. Therefore,

$$T_{OVE} \propto N_A \tag{8}$$

Moreover, as $T_{OVE}$ holds the time consumed by communication, it is also proportional to the number of queens, $N$. It is noticeable that $T_{OVE}$ is not proportional to $N^2$ because the result is received through an one dimensional vector of length $N$.

$$T_{OVE} \propto N \tag{9}$$

Combining (7) and (8)

$$T_{OVE} \propto N_A N \tag{10}$$

or

$$T_{OVE} = K_{OVE} N_A N \tag{11}$$

where $K_{OVE}$ is a proportionality constant that varies network to network.

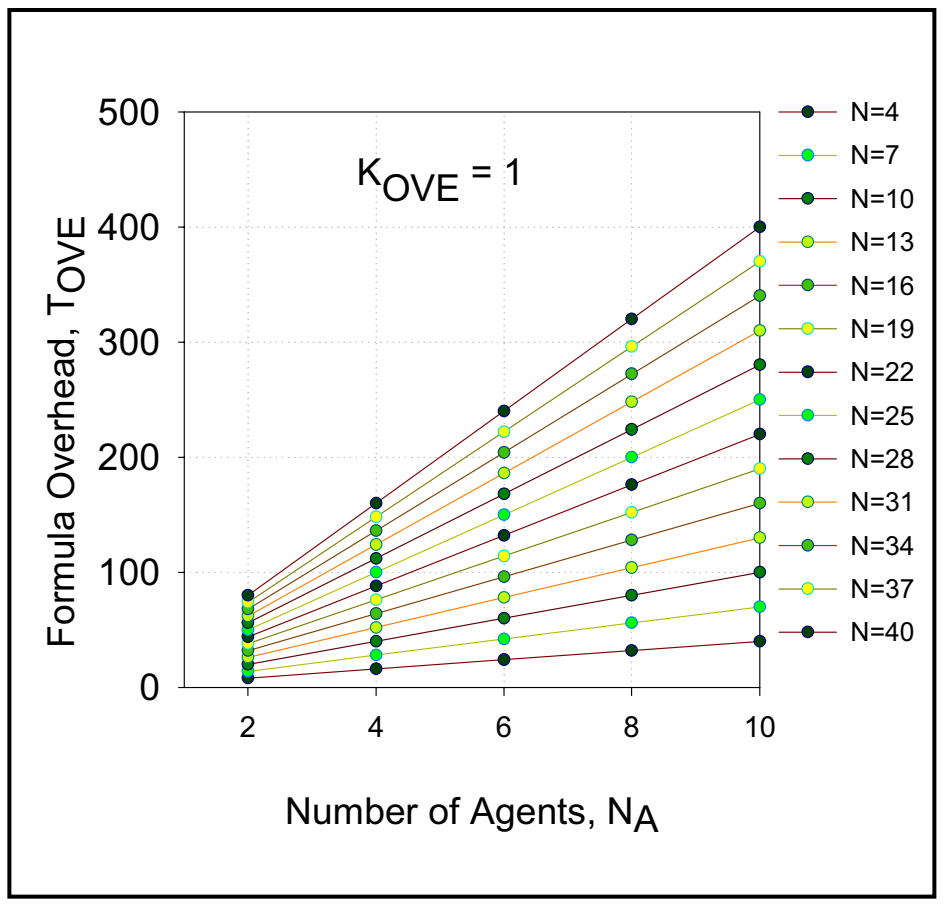

**Figure 12:** Formula Overhead.

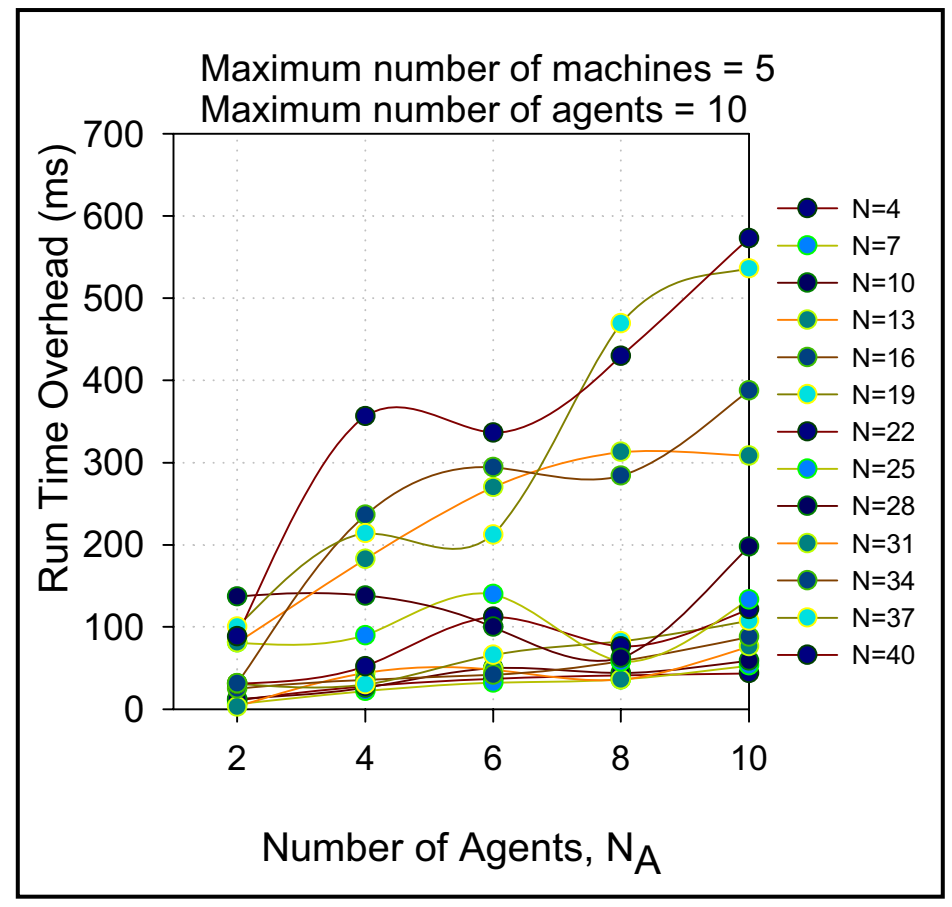

**Figure 13:** Runtime Overhead.

Figure 12 shows the formula overhead plotted by equation (11) and Figure 13 shows the corresponding runtime results. All the points of the runtime plots of this paper are average of multiple number of simulations. If the averages were taken from infinite number of simulations, the resulting plots would become as perfect as the formula graphs.

The most important feature of the analysis should be derived from the ratio $\frac{T_O}{T_{OVE}}$. If it turns to unity i.e., the actual search time becomes equal to the overhead, addition of further agents should be staopped. Because futher involment of agents would hinder the speedup increment.

From equation (7) and equation (11), it is obtained that

$$\frac{T_O}{T_{OVE}} = \frac{\frac{K_O N^2}{N_A}}{K_{OVE} N_A N} = \frac{K_O}{K_{OVE}} \times \frac{N}{(N_A)^2} \tag{12}$$

The ratio $\frac{K_O}{K_{OVE}}$ plays an important role for getting the optimum result and the terminal point of involment of agents. The value, $\frac{N}{(N_A)^2}$ should never be less than $\left[\frac{K_O}{K_{OVE}}\right]^{-1}$, otherwise $T_O$ would become less than $T_{OVE}$.

In this experiment, $K_O = 250$ and $K_{OVE} = 1$. So, the *ultimate number of agents*, when $N = 40$ can be deduced as

$$\frac{K_O}{K_{OVE}} \times \frac{N}{(N_A)^2} = 1, \quad \Rightarrow \frac{250}{1} \times \frac{40}{(N_A)^2} = 1, \quad i.e., N_A = 100.$$

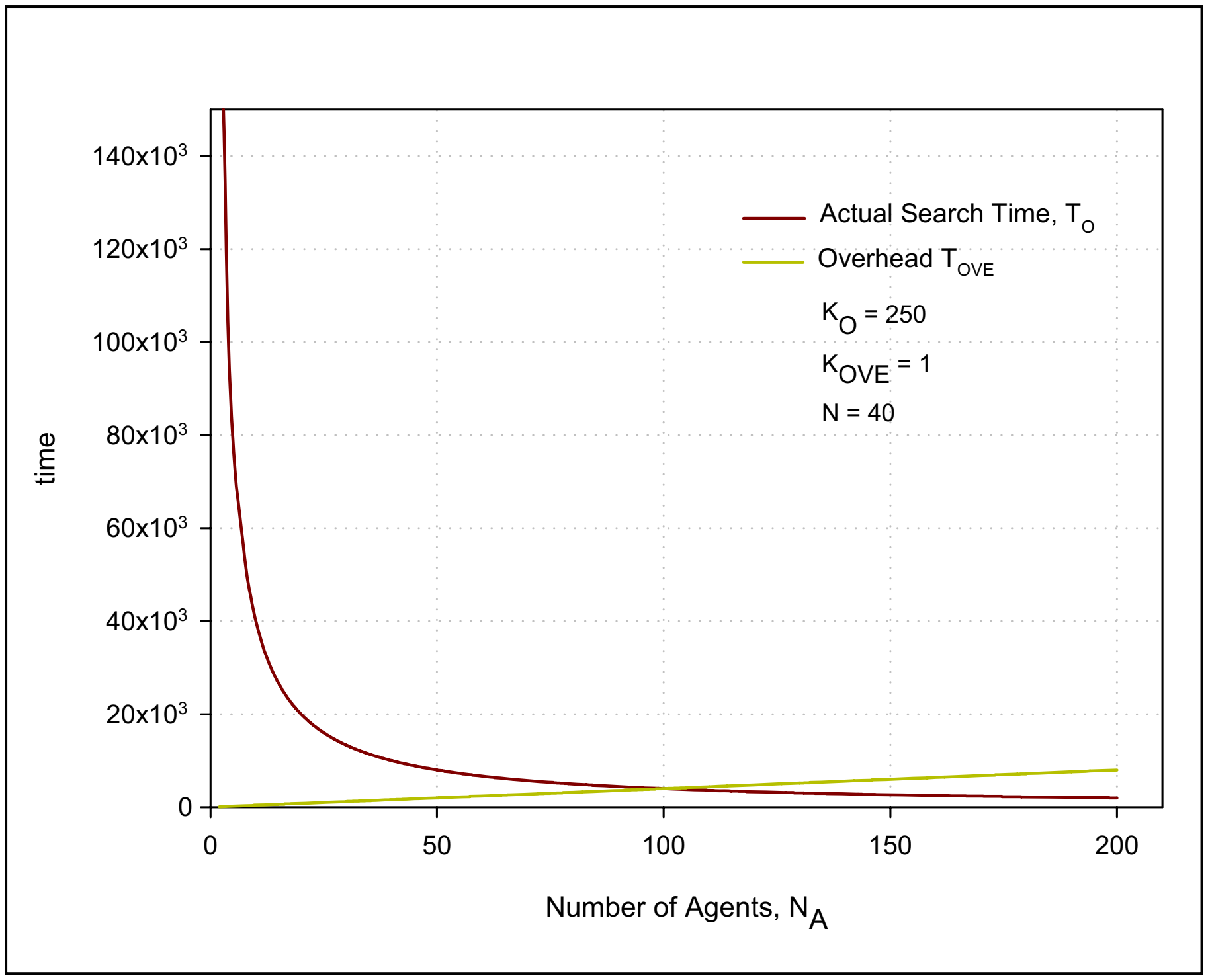

**Figure 14:** Terminal point at $N_A$ =100 while $N$ = 40.

Figure 14 also shows the theoretical plot, which shows that if more than 100 agents are used in a parallel system, the overhead would dominate. So, in a distributed and parallel system for solving *Constraint Satisfaction Problem*s, there exists an *ultimate number of agents*, after which the *Turn Around Time* is dominated by the overhead.

## 4. FUTURE DIRECTION

The future direction of the work is to provide Agent Migration to employ the under utilized machines of the network. Establishment of Agent Migration by *Thread Migration*[14] mechanism would develop a very dynamic system that ensures maximum use of the resources. Migration of agent to an underutilized machine from a loaded one would provide a balance of loads in the Local Area Network.

## 5. CONCLUSION

Global Evaluation Function itself produces better result depending on some probability in searching for solution of Constraint Satisfaction Problems. Moreover, if a distributed mechanism based on *maxAgent* strategy is built, it would definitely provide faster response, as experimental data shows that 70% GEF based randomization provides better results than typical solutions. In a distributed system, GEF utilizes this probability with maximum possible efficiency because even if the probability becomes smallest, the initiator would receive results within the shortest possible time from the machine that is nearest to the solution. Hence *maxAgent* strategy along with *ultimate number of agents* in a concurrent fashion ensures maximum benefit over typical solutions and enriches parallel processing systems for solving *Constraint Satisfaction Problem*s.

## 6. REFERENCES

[ 1 ] Jing H. and Qingsheng C., "Emergence from Local Evaluation Function", *SFI (Santa Fe Institute)*, WP 02-08-036, August 2002.

[ 2 ] Jing H., Liu J. and Cai Q., "Agents to a Kingdom of N Queens", *Jiming Liu and Ning Zhong (Eds.), Intelligent Agent Technology: Systems, Methodologies and Tools*, Page 110-120, The World Scientific Publishing Co. Pte, Ltd., Nov. 1999.

[ 3 ] Liu J., Jing H. and Tang Y. Y., "Multi-agent Oriented Constraint Satisfaction", *Artificial Intelligence*, Vol. 136, No. 1, Page 101-144, 2002.

[ 4 ] Kumar, V. 1987. "Depth-first Search", *Encyclopedia of Artificial Intelligence*, Vol. 2, ed. S. C. Shapiro, Page 1004 -1005. New York: John Wiley and Sons, Inc, 1987.

[ 5 ] Kirkpatrick S., GellatJr C. D. and Veechi M. P., "Optimization by Simulated Annealing", *Science*, Vol. 220, Page 671-681, May 1983.

[ 6 ] Boettcher S. and Percus A. G., "Nature's way of optimizing", *Artificial Intelligence*, Vol 119, Page 275-286, 2000.

[ 7 ] McAllester D., Selman B., and Kautz H., "Evidence for Invariants in Local Search", *Proceedings of AAAI'97*, Page 321-326, 1997.

[ 8 ] Selman B., Henry A. Kautz, and Cohen B., "Noise Strategies for Improving Local Search", *Proceedings of AAAI'94*, Page 337-334. MIT Press, 1994.

[ 9 ] Glover F., "Tabu Search - Part I", *ORSA Journal of Computing*, Vol. 1, No. 3, Page 190-206, 1989.

[ 10 ] Glover F., "Tabu Search - Part II", *ORSA Journal of Computing*, Vol. 2, No. 1, Page 4-32, 1990.

[ 11 ] Vipin Kumar, "Algorithm for Constraint Satisfaction Problem: A Survey", *AI Magazine*, Vol. 13 No. 1, Page 32-44, 1992.

[ 12 ] Sosic R., Gu J., "Efficient local search with conflict minimization: A case study of the N-Queen problem", *IEEE Transactions on Knowledge and Data Engineering*, Vol.6, No.5, Page 661-668, 1994.

[ 13 ] Hossain M. S., Fuad M. M., Deb D., Khan K.M.N.H., and Joarder M. M. A., "Homogenization: A Mechanism for Distributed Processing across a Local Area Network", To appear in the *ECOOP'2004 Workshop on WS 7: Communication Abstractions for Distributed Systems,* Oslo, Norway, June 14, 2004.

[ 14] K. Thitikamol and P. J. Keleher , "Thread Migration, Load Balancing and Heterogeneity in Non-Dedicated Environments"*, Proceedings of the 2000 Int'l Parallel and Distributed Processing Symposium*, Page 583-588, May 2000.